\documentclass[useAMS, usenatbib]{mn2e}

\usepackage[dvips]{graphicx}

\input{epsf}

\usepackage{amssymb}

\usepackage{multirow}

\newcommand{\msunyr}{\mbox{\thinspace $\,M_{\odot}$yr$^{-1}$}}
\newcommand{\ergscm}{\mbox{\thinspace erg\thinspace s$^{-1}$\thinspace cm$^{-2}$}}

\newcommand{\kms}{\thinspace km \thinspace s$^{-1}$}

\newcommand{\ha}{\mbox{H$\alpha$}}

\newcommand{\lya}{\mbox{Ly$\alpha$}}

\newcommand{\OIII}{\mbox{[O\thinspace{\sc iii}]}}

\newcommand{\OII}{\mbox{[O\thinspace{\sc ii}]}}

\newcommand{\kmsmpc}{\thinspace km\thinspace s$^{-1}$\thinspace Mpc$^{-1}$}

\title[\OIII\ emission from DLA galaxies]{The first detection of
\OIII\ emission from high--redshift damped Lyman--$\alpha$
galaxies\thanks{Based on observations made at the European Southern
Observatory Very Large Telescope, Paranal, Chile (ESO Programs
63.O-0618 and 65.O-0707)}} \author[S. J. Weatherley, et al.]
{S.~J.~Weatherley$^1$\thanks{Email :
stephen.weatherley@imperial.ac.uk}, S.~J.~Warren$^1$, P.~M\o ller$^2$,
S.~M.~Fall$^3$, J.~U.~Fynbo$^4$, \newauthor S.~M.~Croom$^{5}$\\
$^1$Astrophysics Group, Blackett Laboratory, Imperial College London,
Prince Consort Road, London SW7 2BW, UK\\ $^2$European Southern
Observatory, Karl-Schwarzschild-Strasse 2, D-85748 Garching bei
M\"{u}nchen, Germany\\ $^3$Space Telescope Science Institute, 3700 San
Martin Drive, Baltimore, MD21218, USA\\ $^4$Institute of Physics and
Astronomy, University of \AA arhus, DK-8000 \AA arhus C, Denmark\\
$^5$The Anglo-Australian Observatory, PO Box 296, Epping, NSW 2121,
Australia}

\date{Accepted 0000 January 00.
      Received 0000 January 00;
      in original form 0000 January 00}

\pagerange{\pageref{firstpage}--\pageref{lastpage}}
\pubyear{2004}
\begin{document}
\maketitle
\label{firstpage}


\begin{abstract}

We present the detection of \OIII\ emission lines from the galaxies
responsible for two high--redshift $z>1.75$ damped Lyman--$\alpha$\
(DLA) absorption lines. We find two sources of \OIII\ emission
corresponding to the $z=1.92$ DLA absorber towards the quasar
Q\,2206$-$1958, and we also detect \OIII\ emission from the galaxy
responsible for the $z=3.10$ DLA absorber towards the quasar
2233.9$+$1381. These are the first detections of rest--frame optical
emission lines from high--redshift DLA galaxies. Unlike the \lya\ line,
the \OIII\ line provides a measure of the systemic velocity of the
galaxy. We compare the \OIII\ redshifts with the velocity profile of
the low--ionisation metal lines in these two absorbers, with the goal
of distinguishing between the model of Prochaska and Wolfe of DLA
absorbers as large rapidly rotating cold thick discs, and the standard
hierarchical CDM model of structure formation, in which DLAs arise in
protogalactic fragments. We find some discrepancies with the
predictions of the former model. Furthermore the image of the DLA
galaxy towards Q\,2206$-$1958 shows a complex disturbed morphology,
which is more in accord with the hierarchical picture. We use the
properties of the rest--frame optical emission lines to further
explore the question posed by M\o ller et al.: are high--redshift DLA galaxies
Lyman--break galaxies (LBGs) selected by gas cross section? The
measured velocity dispersions of the DLA galaxies are in agreement
with this picture, while the data on the \OIII\ luminosities and the
velocity differences between the \lya\ and \OIII\ lines are
inconclusive, as there are insufficient LBG measurements overlapping
in luminosity. Finally we estimate the star formation rates in these
two DLA galaxies, using a variety of diagnostics, and include a
discussion of the extent to which the \OIII\ line is useful for this
purpose.

\end{abstract}




\section{Introduction}

The damped Lyman--$\alpha$ (DLA) absorption lines detected in the spectra of
quasars identify gas clouds that contain the majority of the neutral
hydrogen in the Universe. Analysis of statistical samples of DLA
absorbers has been used to study the cosmic history of star formation
(e.g. Pei, Fall, and Hauser, 1999), and of metal production (Kulkarni
and Fall, 2002). We are engaged in a programme to identify the
galaxies (hereafter `DLA galaxies') responsible for the DLA lines at
high redshifts, $z>1.75$ (Warren et al., 2001; hereafter W01). The
principal goal of this programme is to establish the connection
between the population of DLA absorbers, and galaxy populations
identified at high redshifts in deep imaging studies, firstly by
comparing the measured properties of the detected DLA galaxies with
the properties of other galaxy populations (M\o ller et al., 2002;
hereafter M02), and secondly by measuring the gas cross sections of
the absorbers, thereby establishing their space density (Fynbo, M\o
ller, \& Warren, 1999; Chen \& Lanzetta, 2003).

DLA galaxies are difficult to identify because they typically lie at
angular separations of order one arcsec from the quasar line of sight,
and are therefore swamped by light from the quasar. We are aware of
only eight high--redshift DLA galaxies with published spectroscopic
confirmation of the redshift\footnote{as well as three unpublished
(Djorgovski, private communication)}.
For reference, we have summarised details of these
detections in Table 1. In successive columns are listed (1) the quasar
name; (2) the quasar redshift $z_{QSO}$; (3) the DLA redshift
$z_{DLA}$; (4) a flag Y if $z_{QSO}\gg z_{DLA}$, N otherwise
($z_{DLA}\approx z_{QSO}$); (5) the column density
$\mathrm{log_{10}}N_{\rm HI}$; (6) a flag Y if
$\mathrm{log_{10}}N_{\rm HI}>20.3$ (the definition of DLA of Wolfe et al. 
1986), N otherwise;
(7) the metallicity, if an accurate Si or Zn
value is available; (8) the reference for the metallicity; (9) the
\lya\ luminosity; and (10) the reference of the discovery paper. We
treat all these sources as DLA galaxies, but note that different individuals
adopt narrower definitions of a DLA absorber 
dependent on the flags in columns (4) and (6), so
that under the narrowest definition only three sources are DLA
galaxies. For a discussion of the cases where $z_{DLA}\approx
z_{QSO}$, see M\o ller, Warren, and Fynbo (1998). 
Five of the DLA galaxies listed in Table 1 were discovered
with telescopes of 4m--class, or smaller (the remaining three are from
the current programme). 
Motivated by the high spatial resolution of
{\it HST}, and the large light--gathering power of 8m--class telescopes, we
are undertaking a survey to detect a significant sample of DLA
galaxies (W01). We have obtained deep {\it HST} NICMOS and STIS images of
the fields of 16 quasars, aimed at the detection of counterpart
galaxies of 18 $z>1.75$ DLA absorbers and five Lyman--limit systems. In
the NICMOS images we found 41 candidate DLA galaxies brighter than
$H_{AB}\sim 25$, within a box of side $7.5$\thinspace arcsec centred
on each quasar (W01). We are using the VLT and Gemini telescopes to
obtain confirmatory optical and near--ir spectra of these
candidates.

\begin{table*}
\caption{Summary of spectroscopically confirmed high--redshift DLA galaxies}
\label{table:DLAs}
\centering
\begin{scriptsize}
\begin{tabular}{lllclccccl}\\
\hline

 1 & 2 & 3 & 4 & 5 & 6 & 7 & 8 & 9 & 10 \\
Quasar & $z_{\rm QSO}$ & $z_{\rm DLA}$ & $z_{\rm QSO} \gg z_{\rm DLA}$ & 
$\mathrm{log_{10}}(N_{\rm HI}$) & $\mathrm{log_{10}}(N_{\rm HI}>20.3$) &
 [M/H]  & M (refs) & Ly$\alpha$ Lum  & Original \\
 & & & & cm$^{-2}$ & cm$^{-2}$ &  &  & $\times 10^{42}$ erg\,s$^{-1}$ & Discovery \\

\hline

PHL\,1222        & 1.922 & 1.9342 &  N  & 20.36 &  Y  & --- & --- &
$\sim 9.3$ & [12] \\

PKS\,0458$-$02   & 2.286 & 2.0395 &  Y  & 21.65 &  Y  & $-1.17,-1.19$ & Zn,Zn $^{\rm [5],[14]}$ &  
$1.6^{+0.6}_{-0.2}$ & [9] \\

PKS\,0528$-$250  & 2.797 & 2.8110 &  N  & 21.35 &  Y  & $-0.75,-0.76$ & Si,Zn $^{\rm [7],[5]}$ & 
$5.2\pm0.4$  & [10] \\

PC\,0953$+$4749  & 4.457 & 3.407  &  Y  & 21.2  &  Y  & $> -2.09$ & Si$^{\rm [15]}$ &
$\sim1.1$ & [1] \\

Q\,2059$-$360    & 3.097 & 3.0825 &  N  & 20.85 &  Y  & --- & --- &
$\sim17$ $^{[a,6]}$ & [13] \\

Q\,2206$-$1958   & 2.559 & 1.9205 &  Y  & 20.65 &  Y  & $-0.42,-0.39$ & Si, Zn $^{\rm [14],[5]}$ & 
$6.8\pm0.8$ $^{\rm [b]}$  & [11] \\

2233.9$+$1318    & 3.298 & 3.1501 &  Y  & 20.00 &  N  & $ -1.04\,{\rm to}\,-0.56$ & Si $^{\rm [c,8]}$ & 
$5.6\pm1.0$ & [2] \\

DMS\,2247$-$0209 & 4.36  & 4.097  & Y & ---   & --- & --- & --- & 
$0.9\pm0.2$ $^{\rm [4]}$ & [3] \\

\hline
\end{tabular}
\begin{minipage}{180mm}
Notes: [a] assumes a total flux $2.0\times10^{-16}$
erg\,s$^{-1}$\,cm$^{-2}$, [b] this is the luminosity of N-14-1C, [c] lower and upper limits provided by ref.[8]. All
values in columns 2, 3, 5 are taken from the summary of Warren et
al. (2001), except for Q\,$2059-360$ (taken from [7]),
and for DMS\,$2247-0209$ (taken from [3]).
The fluxes used to compute the luminosities are taken from the
discovery paper, unless referenced otherwise. References are as
follows: 
[1] Bunker et al. (2005, in prep.),
[2] Djorgovski et al. (1996),
[3] Djorgovski et al. (1998),
[4] Djorgovski (private communication), 
[5] Kulkarni and Fall (2002), 
[6] Leibundgut and Robertson (1999), 
[7] Lu et al. (1996), 
[8] Lu et al. (1998), 
[9] M\o ller, Fynbo \& Fall (2004),
[10] M\o ller \& Warren (1993),
[11] M\o ller et al. (2002),
[12] M\o ller, Warren \& Fynbo (1998),
[13] Pettini et al. (1995),     
[14] Prochaska et al. (2003a), 
[15] Prochaska et al. (2003b). 
\end{minipage}
\end{scriptsize}
\medskip
\end{table*}

In M02, we reported preliminary results from this programme. We tested
the hypothesis that DLA galaxies are Lyman--break galaxies, selected by
gas cross section, by comparing several emission properties (size,
colour, etc.) of three high--redshift DLA galaxies with the emission
properties of Lyman--break galaxies of similar absolute magnitude and
redshift. We found no significant differences and concluded that the
data are consistent with the hypothesis posed.  It must be
appreciated, though, that the hypothesis that the two populations are
drawn from the same parent population, with different selection
criteria, allows for the possibility that the average properties of
the population (e.g. their clustering amplitude) will differ. This
will be true, for example, for any quantity that depends on
luminosity, since the average luminosities of the two populations
differ, due to the way they are selected. As noted by M02, the key to
confirming the hypothesis conclusively is the detection of more DLA
galaxies, which will allow the measurement of the relation between gas
radius and galaxy luminosity, the Holmberg relation $R_{gas}\propto
L^{\rm t}$. The Holmberg relation provides the transformation between
the luminosity distributions of the two populations (see Fynbo et
al. 1999 for further explanation).

In this paper we report the detection of rest--frame optical emission
lines from two DLA galaxies, with near--ir spectroscopy. These are the
first detections of this kind\footnote{The detection of H$\beta$ and
\OII\ reported by Elston et al., 1991, was not confirmed (Lowenthal, private communication).}. The
galaxies are two of the three studied in M02 (for the third, all the
strong rest--frame optical emission lines lie at highly unfavourable
wavelengths for observation). We use these results to extend our
comparison of the properties of DLA galaxies and LBGs (of similar
absolute magnitude and redshift) to include the rest--frame optical
line luminosities and widths.

In the CDM scheme for the formation of structure in the Universe,
galaxies grow hierarchically, and DLA absorption lines at $z\sim2-3$
arise when sightlines pass through protogalactic fragments. Prochaska
and Wolfe (1997b) have argued for a very different picture, showing
that that the detailed velocity structure of the low--ionisation metal
absorption lines in DLA systems is consistent with the expectation for
sightlines passing through large rapidly--rotating cold thick discs of
neutral gas. However, it was subsequently shown by Haehnelt,
Steinmetz, and Rauch (1998) that the absorption--line kinematic data
alone do not permit an unambiguous interpretation, and are equally
well explained by the dynamics of merging protogalactic fragments.
Our detections of rest--frame optical emission lines from DLA galaxies
provide important additional clues for the interpretation of the
kinematics of these systems, supplementing the information provided by
the absorption--line velocity profiles, and casting new light on this
long--standing debate. In the same context we note that additional
evidence in favour of the hierarchical picture comes from recent
observations of the evolution of the sizes of galactic stellar disks
(Ferguson et al., 2004, Bouwens et al., 2004), and comparison against
predictions of the extent of the baryons in galaxies, as a function of
redshift (Fall and Efstathiou, 1980, Mo, Mao, and White, 1998). All
the same these analyses rely on assumptions about the extent to which
stars map the baryons at any redshift. The analysis of the kinematics
of DLA absorbers complements this approach.

The layout of the paper is as follows: in Section 2 we describe the
observations and the data reduction; in Section 3 we present the
measured properties of the detected lines; in Section 4 we analyse the
kinematics of these systems, and in Section 5 we compare the measured
properties of the detected galaxies against the properties of LBGs.
In section 6 we estimate the star formation rates in these two DLA
galaxies, using a variety of diagnostics. Section 7 provides a
summary of the main results of the paper. Throughout, we assume a
standard, flat $\Lambda$CDM cosmology with $\Omega_\Lambda = 0.7$ and
H$_0 = 70$\kmsmpc.  For this cosmology an angle of 1 arcsec
corresponds to the physical size 8.4 kpc at $z=2$, and 7.7 kpc at
$z=3$.


\section{Observations and data reduction}
\label{sec:observations}

\subsection{Observations}
The observations were taken with the Infrared Spectrometer and Array
Camera (ISAAC) instrument on the European Southern Observatory's 8m
UT1 telescope at the Very Large Telescope (ESO--VLT). We used the medium--resolution (MR)
mode, with a $1$\thinspace arcsec slit, to obtain near--ir spectra of a
number of targets in the NICMOS candidate list from W01. Results for
three candidates are reported here. Results for other candidates will
be reported elsewhere.  We observed two candidates in the field of the
quasar Q\,2206$-$1958, in the H band, where the MR mode provides a
wavelength coverage of $0.079\mu$m, with 1024 pixels, at a resolving
power of 2700 for this slit, corresponding to 7 pixels. The third
target was the known DLA galaxy in the field of the quasar
2233.9$+$1318 (Djorgovski et al., 1996, and see Table 1), which was
observed in the K band, where the MR mode provides a wavelength
coverage of $0.122\mu$m, at a resolving power of 2750, corresponding
to 7\thinspace pixels. The spatial scale is $0.147$\thinspace arcsec per
pixel. For spectroscopy in the near--ir it is common practice to nod
the telescope between two positions $A$ and $B$, in sequence $ABBA$
etc.  First--order sky subtraction is achieved by subtracting the
average spectrum at position $B$ from the average spectrum at position
$A$, and {\em vice versa}. Instead, we placed the galaxy at six
positions along the slit, $ABCDEF$, each separated by 7\thinspace
arcsec. For each slit position, first--order sky subtraction at any
position is achieved by subtracting the mean of the frames at all the
other sky positions. For observations of faint targets, this procedure
should lead to an increase in signal--to--noise (hereafter $S/N$) of a
factor $\sqrt{2(N-1)/N}$ where $N$ is the number of slit--positions;
for six slit--positions we expect an increase in $S/N$ of $\sim1.3$
compared to the normal procedure.

The journal of observations is provided in Table 2. Columns 1 and 2
list the names of the quasar, and of the target DLA galaxy
candidate. The target names are taken from W01, and may be decoded
from the example N-14-2C: N=NICMOS candidate, 14=14th quasar in the
list of 16 quasars targeted with NICMOS, 2=2nd nearest candidate to
the line of sight to the quasar, C=compact morphology (as opposed to
D=diffuse). As detailed in W01, in this context the meaning of
`compact' is that the $S/N$ of the detection, integrated over an
aperture, is greater for the smaller aperture used, diameter
$0.45$\thinspace arcsec, than for the larger aperture used, diameter
$0.90$\thinspace arcsec. Successive columns in Table 2 list the date
of observation, the total integration time, the wavelength range
covered by the spectrum, the slit position angle, and the average
seeing.  The observations of 2000 Aug 2 were affected by cloud, while
conditions were clear for the other two nights. Further details of the
target DLA absorbers, and the candidate DLA galaxies, including
accurate coordinates, are provided in W01. Notes on the three
candidates follow.

\begin{figure}
\begin{center}
\includegraphics[width=7cm]{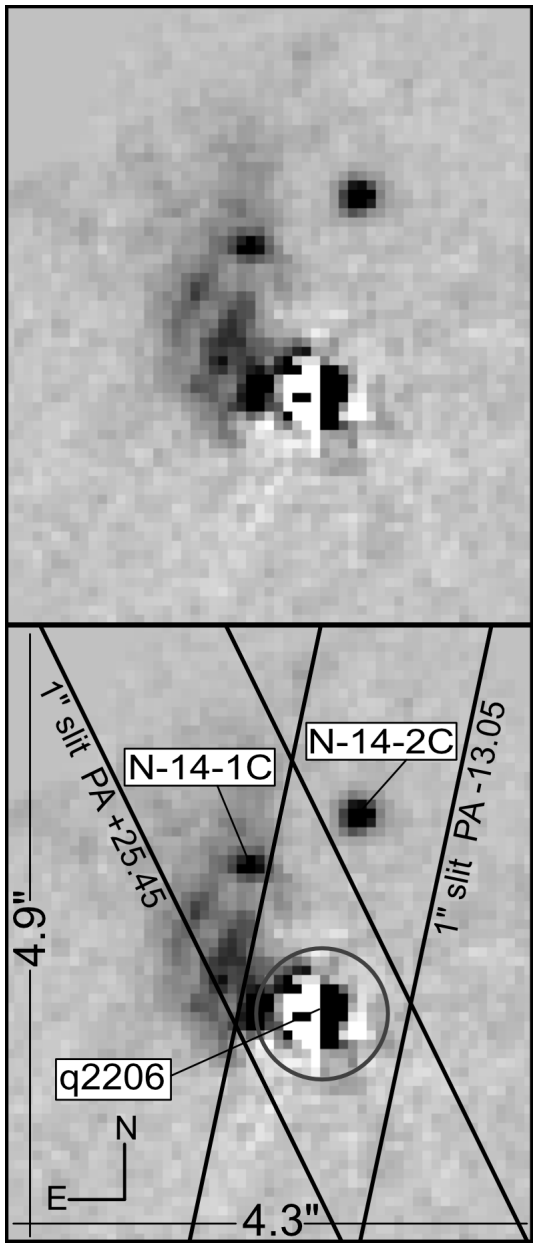}
\caption{ \label{fig:2206_field_final} Sub--section of the STIS 50CCD image of the field toward
Q\,2206$-$1958 after quasar {\em psf} subtraction. The upper image is
the same as the lower image, but without the annotation. N is up and E
to the left. The smooth region in the upper LH corner contains no data
in this image, because the image was formed from a small region of the
STIS field, and then rotated to orient it N--S. The two candidates
N-14-1C and N-14-2C, as well as the location of the quasar, are
indicated by arrows. The ring centred on the location of the quasar
marks a radius of $0.4$\thinspace arcsec. The {\em psf} subtraction is
considered to be satisfactory outside this radius, and unsatisfactory
inside. Details of the {\em psf} subtraction procedure used will be
provided elsewhere (M\o ller et al., in prep.). The sky coverage of the
$1$\thinspace arcsec slit in the two orientations is shown. The pixel
size is $0.056$\thinspace arcsec. The pixel size in the original STIS
images is $0.050$\thinspace arcsec. The image is a combination of four
images taken at two different spacecraft roll angles, totaling 5084s
integration. An approximate $5\sigma$ depth of this image, for a point
source, well away from the quasar centroid is $V_{50}=27.65$. }
\end{center}
\end{figure}

{\em Quasar Q\,2206$-$1958, candidates N-14-1C, N-14-2C:\,} The
spectrum of this quasar shows two high--redshift DLA absorbers, of
redshifts $z=1.9205$, and $z=2.0762$. Our NICMOS observations of this
field revealed two compact candidates with impact parameters
$b<2.0$\thinspace arcsec. The nearest candidate to the line of sight
to the quasar, N-14-1C, has impact parameter $b=1.13$\thinspace
arcsec, and was detected at $S/N=6.5$. Nevertheless this source was
considered a marginal detection since the image is located on a
diffraction spike. The second nearest candidate, N-14-2C, has impact
parameter $b=1.33$\thinspace arcsec, and was detected at $S/N=12.1$.
As reported in M02, we succeeded in detecting Ly$\alpha$ emission from
N-14-1C, confirming that it lies at the redshift of the DLA absorber
at $z=1.9205$. We also observed N-14-2C, but failed to secure a
redshift. The source is redder than N-14-1C, is unresolved in the STIS
and NICMOS images, and we assumed it was unrelated\footnote{Because
of the wider slit used, $1.3$\thinspace arcsec, and the poor seeing conditions,
although weak \lya\ emission was detected in the spectrum of N-14-2C,
the flux was consistent with spillage from N-14-1C.}. In this paper we
show that, in fact, the two sources are at the same redshift.

The STIS image of this field, provided in M02, detects both candidates
at much higher $S/N$, by about a factor of five.  This image was taken
without a filter i.e. in `50CCD' mode.  Another version of this image
is reproduced here in Fig. 1, rotated so that N is up (and E to the
left). The deeper STIS image reveals a more complex morphology than the NICMOS image. In addition
to the two compact sources, some diffuse emission surrounds N-14-1C,
and extends to smaller impact parameters.

For the field of the quasar Q\,2206$-$1958, the choice of central
wavelength for the two observations attempted to maximise the chances
of successful detections, from consideration of several factors: the
wavelengths and probable relative strengths of the various rest--frame
optical emission lines at the redshifts of the two absorbers; the
wavelength regions of strong atmospheric absorption; the results from
the optical spectroscopy; the wavelength coverage of the instrument;
and the possibility of other galaxies undetected in the imaging
observations, because at very small impact parameter. For both
observations the slit was centred on the quasar and rotated to the
position angle of the candidate. We chose the same wavelength range
for both slit orientations, which provided simultaneous coverage of
\OIII\,500.7 at $z=1.9205$ (at $1.462\mu$m), and H$\beta$ and
\OIII\,495.9 at $z=2.0762$ (at $1.495$ and $1.525\mu$m). The purpose
of the observation at PA$+$25.45, then, was to detect the \OIII\,500.7
line from the confirmed galaxy N-14-1C, while also allowing the
possibility of detecting the galaxy counterpart to the $z=2.0762$
absorber if it lies at very small impact parameter. The purpose of the
observation at PA$-$13.05, was to search for \OIII\ emission at either
redshift from N-14-2C, as well as to search again for the $z=2.0762$
galaxy at very small impact parameter. It would have been possible to
observe both targets with a single slit orientation, offset from the
quasar; however, we used two slit orientations as we considered it important to
check for emission at very small impact parameter. In addition, in
trying to detect faint emission lines from a galaxy where the spectrum
overlaps spatially with the quasar spectrum, and where the quasar
spectrum must be subtracted (see below), our experience is that this
works best when the slit is aligned along the line joining the quasar
and the galaxy.

{\em Quasar 2233.9$+$1381, candidate N-16-1D:\,} The spectrum of this
quasar shows a strong \lya\ absorption line, $\mathrm{log_{10}}(N_{HI})=20.0$, at
$z=3.1501$. By the DLA definition of Wolfe et al. (1986),
$\mathrm{log_{10}}(N_{HI})>20.3$, this is not a DLA absorber, but a
Lyman--limit system. We have included it in Table 1, since the
measured column density lies only a little below this threshold. In
this paper we treat the counterpart galaxy as representative of DLA
galaxies. Nevertheless it should be borne in mind that the ionisation
state of the gas in the absorber is dependent on column density (e.g. Viegas, 1995),
and in this respect this absorber lies at one end of the distribution
of the absorbers listed in Table 1.
The nearest NICMOS candidate N-16-1D, was first detected in an
optical image and identified as a candidate counterpart to the
absorber by Steidel, Pettini, and Hamilton (1995), who called it
N1. The galaxy was confirmed as the counterpart, on the basis of the
detection of \lya\ emission at the DLA absorber redshift, by
Djorgovski et al. (1996). We did not detect any candidates at smaller
impact parameter in either our NICMOS image (W01) or our STIS image
(M02) of this field. The wavelength range of our observations covered
the three redshifted lines H$\beta$486.1 (at $2.017\mu$m),
\OIII\,495.9 (at $2.058\mu$m), and \OIII\,500.7 (at $2.078\mu$m). The
impact parameter of this galaxy, $2.78$\thinspace arcsec in our NICMOS
image, is the largest of all the eight confirmed galaxy counterparts
listed in Table 1. We aligned the slit to cover both the quasar and
the galaxy, again to check for the possibility that another galaxy at
the absorber redshift lies at very small impact parameter.


\subsection{Data reduction}

For each source, for first--order sky subtraction we followed the
procedure explained in the previous section. If the six--step--nod was
repeated, the frames were reduced in sets of six. For six frames taken
at positions ABCDEF, from each flat--fielded frame we subtracted the average of the
other five frames. Second-order sky subtraction was achieved by
fitting a polynomial up each column. The sky--subtracted frames were
then registered to the nearest pixel, spatially and spectrally, and
combined using inverse--variance weighting. Integer pixel shifts were
used in order to keep the data in each pixel statistically
indepedent. As a check of our six--step--nod observing procedure, we
also reduced the data in the traditional way, by combining pairs of
nod positions. This verified that the six--step--nod procedure achieved
the expected improvement in $S/N$. To establish the uncertainties, for
each final 2D spectrum we formed a corresponding variance frame in two
ways. In the first method we registered the raw frames using the same
offsets as for the sky--subtracted frames. We then summed the counts
at each pixel, and computed the variance as appropriate, from a
knowledge of the measured gain and read noise, and assuming Poisson
statistics. The second method measured the dispersion in the counts up
each column in the final combined sky--subtracted frame, using a
robust estimator. This provides an accurate measure of the average
noise in the sky, but gives no information on the noise for individual
pixels. The second method gave results some 20 to 30\thinspace
per\thinspace cent higher for the variance (10 to 15\thinspace
per\thinspace cent for the standard deviation) than the Poisson
estimate; therefore we scaled the Poisson--estimated variance frame by
this correction factor to produce the final variance frame.

The combined frame formed from the registered raw frames, referred to
above, emulates the registration of the sky--subtracted frames, and
therefore provided the means of accurate wavelength calibration, using
the sky OH emission lines. This frame also provided the means of
measuring the spectral resolution, from the widths of the sky lines,
and yielded the values quoted above.

\begin{table*}
\begin{scriptsize}
\caption{Observation log for spectroscopic observations at VLT}
\label{table:journal}
\centering
\begin{tabular}{lclcclc}\\ 
\hline
\multicolumn{1}{c}{Quasar} & DLA candidate & \multicolumn{1}{c}{Date} 
& Integration & wavelength range & slit PA & seeing \\
       & (W01)         &      & time s     & $\mu$m  & \degr E of N 
& arcsec \\
\hline
Q\,2206$-$199 & N-14-1C  & 2000 Aug 2  & 3600 & $1.453-1.532$ &  $+$25.45 & 1.0 \\
              &          &             &$=6\times600$& & & \\
Q\,2206$-$199 & N-14-2C  & 2000 Jul 8  & 7200 & $1.453-1.532$ &  $-$13.05 & 0.8 \\
              &          &             &$=2\times6\times600$& & & \\
2233.9$+$1318 & N-16-1D  & 2000 Jul 11 & 7200 & $1.988-2.110$ & $+$158.45 & 0.8 \\
              &          &             &$=2\times6\times2\times300$& &           & \\
\hline
\end{tabular}
\end{scriptsize}
\medskip
\end{table*}


\section{Results}

\label{sec:results}

\subsection{Reduced 2D frames}

\begin{figure}
\includegraphics[width=1\columnwidth]{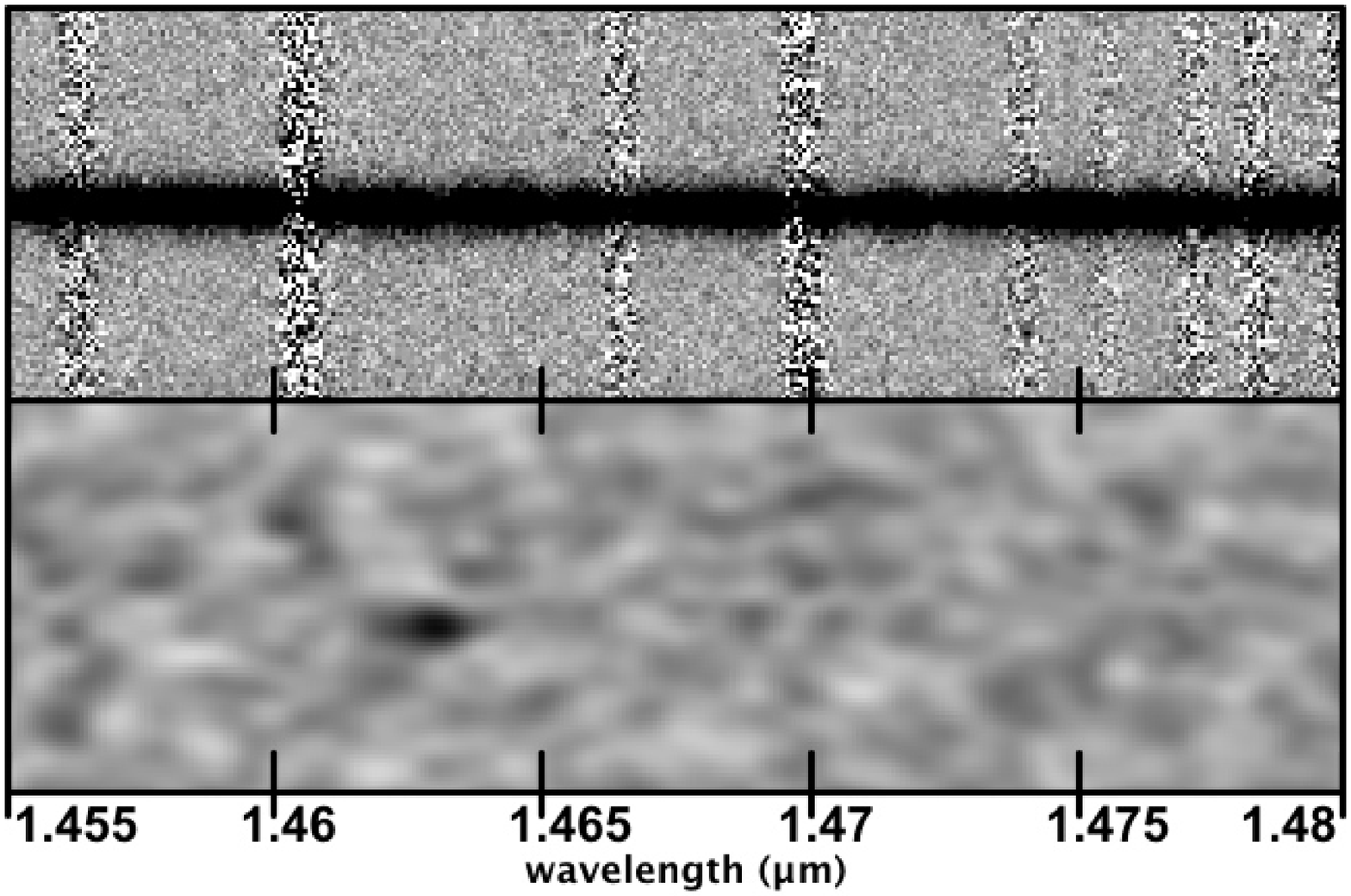}
\includegraphics[width=1\columnwidth]{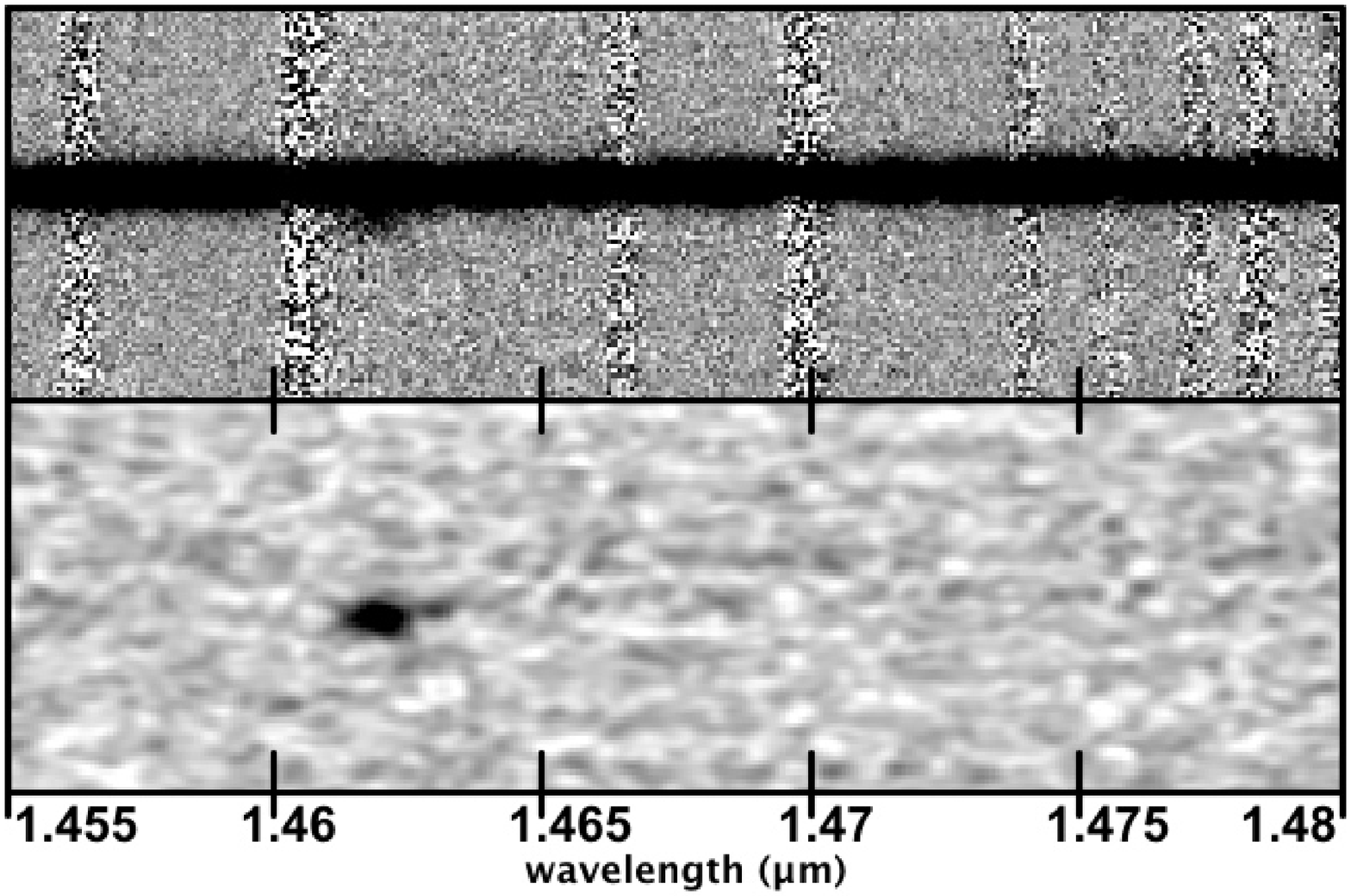}
\caption{ \label{fig:2206_2d} Two--dimensional spectra of Q\,2206$-$1958 both before and after
quasar removal. The vertical bands of higher noise mark the
wavelengths of strong OH sky lines.  From top to bottom: a) final
combined 2D spectrum of observation of candidate N-14-1C, showing the
quasar spectrum. b) the same, after subtraction of the quasar spectrum
using our {\sc SPSF} method, and smoothing, revealing the emission line from
the DLA galaxy N-14-1C. c) final combined 2D spectrum of observation
of candidate N-14-2C, showing the quasar spectrum. The galaxy emission
line is visible in this frame, as an extension below the quasar
spectrum. d) the same, after subtraction of the quasar spectrum using
our {\sc SPSF} method, and smoothing, more clearly revealing the emission
line from the DLA galaxy N-14-2C. The wavelength scale is the same for
all the frames. The velocity offset between N-14-1C in frame b) and
N-14-2C in frame d) is clearly visible. The emission line visible in
frame d) displays a faint but significant extension to redder
wavelengths and smaller impact parameters. The wavelength of this
extension matches the wavelength of the emission line from N-14-1C
visible in b), and is interpreted as light leaking into the slit from
this source. This would be expected, as may be appreciated from
inspection of Fig. 1.}
\end{figure}

{\em Quasar Q\,2206$-$1958, candidates N-14-1C, N-14-2C:\,} The final
2D frames of the spectroscopic observations of Q\,2206$-$1958 are
shown in Fig. \ref{fig:2206_2d}. 
Because of the poorer seeing, shorter integration
time, and non--photometric conditions, the $S/N$ of the spectrum of
N-14-1C (top two panels in Fig. 2) is nearly a factor of two lower
than for N-14-2C (bottom two panels). An emission line was detected at
high significance from both candidates, with centroid below the quasar
spectrum in each frame. For both candidates the impact parameter is
small, and the quasar and galaxy spectra overlap. Therefore we used
our {\sc SPSF} software (M\o ller, 2000) to subtract the quasar spectrum, as
described below. Fig. \ref{fig:2206_2d} provides the 2D spectra over approximately
half the total wavelength range observed, for both candidates, both
before and after subtraction of the quasar spectrum. From top to
bottom, the panels show a) N-14-1C, raw, b) N-14-1C, after quasar
subtraction, and smoothing to enhance the contrast of the line, c)
N-14-2C, raw, d) N-14-2C, after quasar subtraction and smoothing. The
measured spatial offsets of the emission lines from the quasar are
provided in Table 3, together with the offsets measured in the NICMOS
and STIS
images. The spectroscopic and imaging offsets are consistent,
confirming that the emission is associated with the targeted
candidates. The {\sc SPSF} routine works by firstly determining a
normalised profile of the quasar, smoothly varying with
wavelength. This profile is then scaled to the quasar counts at each
wavelength, and subtracted. Regions where emission lines from the
galaxy are detected are then masked, and the procedure is
iterated. This worked well here because the galaxy continuum is
negligibly faint, and because the emission is offset from the quasar
centroid.

\begin{table*}
\begin{scriptsize}
\caption{Spectroscopic and photometric parameters of the three candidates}
\label{table:results1}
\centering
\begin{tabular}{lllllc}\\ 
\hline
Candidate & \multicolumn{3}{c}{impact parameter (arcsec)}& $R$ (spec.)   & $r_{1/2}$ (M02)\\
          & spec.              & image (NIC) & image (STIS)      & (kpc) & (kpc)     \\
\hline
N-14-1C & $0.97\pm 0.04$ & $1.13\pm 0.07$ & $0.99\pm0.05$ & $8.2\pm0.3$ & 4.2 \\
N-14-2C & $1.24\pm 0.03$ & $1.33\pm 0.04$ & $1.23\pm0.05$ & $10.4\pm0.3$ & ---\thinspace $^{\ast}$ \\
N-16-1D & $2.51\pm 0.04$ & $2.78\pm 0.07$ & $2.51\pm0.05$ & $19.0\pm0.3$ & 1.1 \\
\hline
\end{tabular}
\begin{minipage}{100mm}

$^\ast$ consistent with point source

\end{minipage}
\medskip
\end{scriptsize}
\end{table*}

For the candidate N-14-1C the detected line occurs at the expected
wavelength for \OIII\,500.7 from the confirmed galaxy counterpart to
the DLA absorber at redshift $z=1.9205$. For the candidate N-14-2C the
detected line occurs at a very similar, but slightly shorter,
wavelength, confirming that N-14-2C is part of the same DLA galaxy as
N-14-1C, and not an unrelated source as was previously suspected.  A
faint, nevertheless significant, extension of the detected emission
line, towards redder wavelengths and smaller impact parameters, is
visible in the bottom panel in Fig. 2. As explained in the next
sub-section, we interpret this as flux from N-14-1C leaking into the
slit.

\begin{figure}
\includegraphics[width=1\columnwidth]{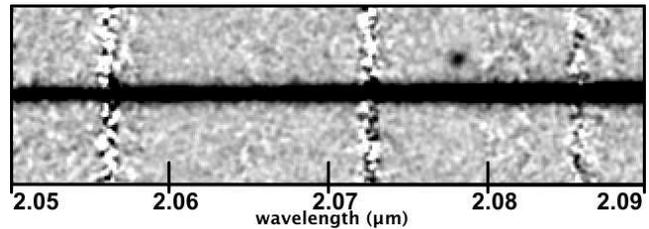}
\caption{ \label{fig:2233_2d} Two-dimensional spectrum of 2233.9$+$1381, smoothed to enhance
the contrast of the detected emission line. The thick line is the
spectrum of the quasar. The \OIII\ emission line from N-16-1D is
clearly visible at a wavelength $2.078\mu$m, offset from the quasar by
$2.5$\thinspace arcsec}
\end{figure}

{\em Quasar 2233.9$+$1381, candidate N-16-1D:\,} Fig. \ref{fig:2233_2d} provides the
2D spectrum of our observation of the quasar 2233.9$+$1381 over
approximately half the total wavelength range observed.  An emssion
line is clearly visible, at the expected wavelength for \OIII\,500.7
from the confirmed DLA galaxy counterpart to the Lyman--limit absorber
at redshift $z=3.1501$. The frame has been smoothed to enhance the
contrast of the line. Neither the \OIII\,495.9 nor the H$\beta$486.1
lines were detected. The detected emission line is well separated from
the quasar, so that it was not necessary to subtract the spectrum of
the quasar. The spatial offset measured from the 2D spectrum, listed
in Table 3, is somewhat smaller than the value measured from the
NICMOS image, at the 3$\sigma$ significance level, while it is in
agreement with the offset measured in the STIS frame. Such a discrepancy might be explained, for example, by
variations in mean age of the stellar populations across the
galaxy. It is interesting to note that the rest--frame UV continuum seems
to trace the \OIII\ line emission better than the rest frame optical continuum.


\subsection{Extracted 1D spectra}

The quasar 1D spectra and the corresponding variance spectra were
extracted using the {\sc IRAF} routine {\it apall}, with a 14--pixel
aperture and an optimal profile weighting scheme. The same apertures,
with appropriate vertical shift, were then applied to the extraction
of the galaxy spectra, and the corresponding variance spectra, from
the 2D frames with the quasar spectra subtracted. At this stage ww
checked the variance spectra, by forming, for each galaxy, the ratio
of the galaxy spectrum and the standard--deviation spectrum. As
expected the counts in these $S/N$ spectra have mean zero, and standard
deviation unity. From the frames formed by summing the registered raw
frames, we re--extracted the quasar spectra, which now include
sky. Vacuum wavelengths of sky emission lines in these spectra were
used to calibrate the quasar and galaxy spectra onto a linear
wavelength scale. All measured wavelengths quoted in this paper have
been corrected to the heliocentric frame.

We used the spectra of bright A{\sc II} and G{\sc IV} stars, observed on the same
nights, at similar airmasses to the targets, to flux calibrate the 1D
spectra. The calibration curve was derived by taking the ratio of
the observed spectrum of each standard, and a black--body spectrum of the
correct brightness. The spectra of Q\,2206$-$1958 are affected by
atmospheric absorption bands, which are corrected for by this
procedure. Neither of the detected emission lines is strongly affected
by absorption. This procedure provides reasonably accurate
calibration provided slit losses for the object and standard star are
similar. Since our targets have angular size much smaller than the
seeing this will be true if the seeing conditions were similar, which
was true for the two nights which were clear. For the data taken on
2000 Aug 2 (N-14-1C), when conditions were not photometric, the
calibration was derived from the ratio of the uncalibrated spectrum of
the quasar from that night, to the calibrated spectrum of the same
quasar observed on 2000 Jul 8 (N-14-2C). As an additional check we
compared the calibrated fluxes of the quasars against the NICMOS
photometry, finding agreement at the 0.1\thinspace mag. level. This check
provides only an indication of the spectrophotometric accuracy,
because the NICMOS and spectroscopic observations were separated by
two years, and the quasars may have varied in the interim.

\begin{figure*}
\includegraphics[width=14cm]{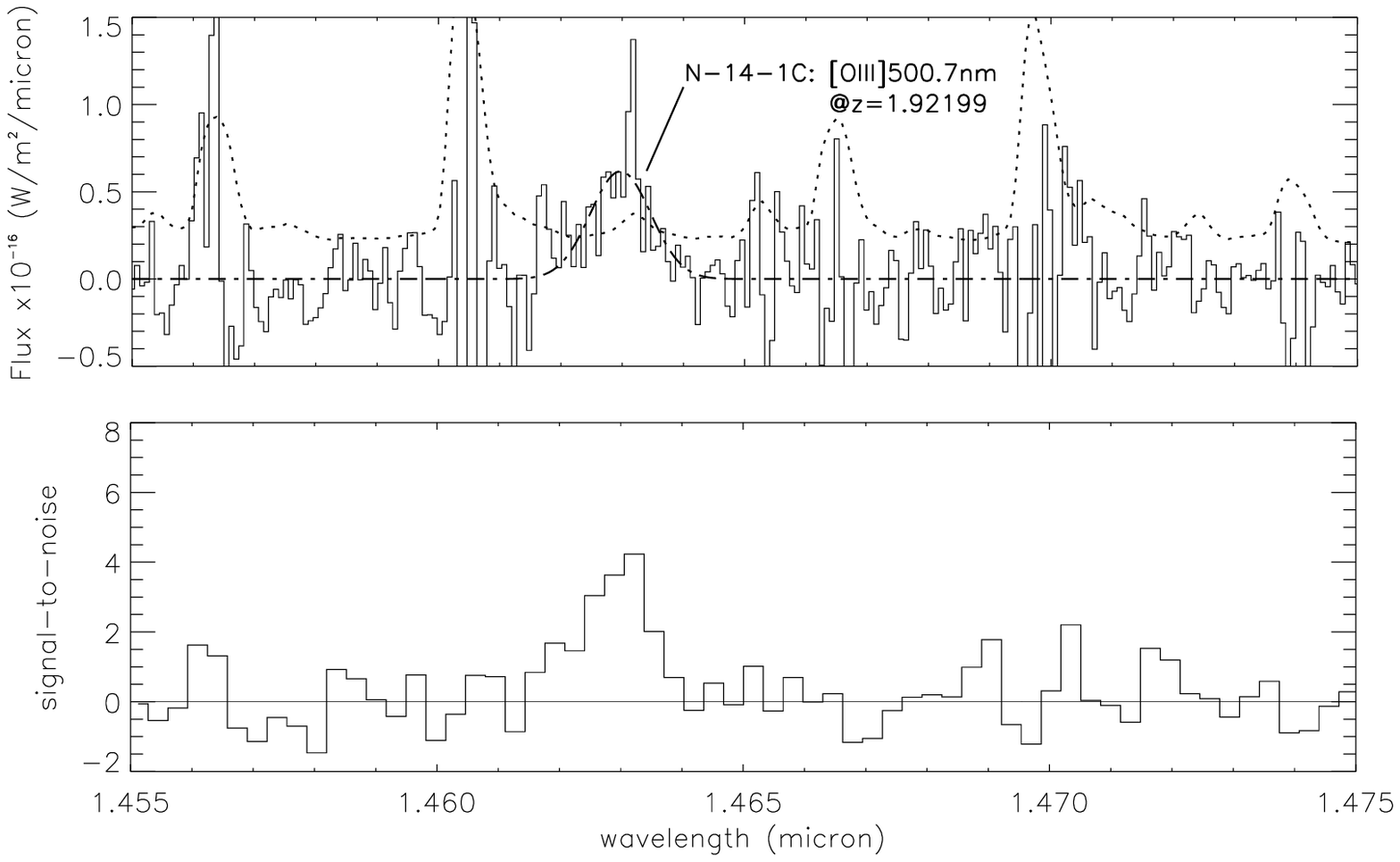}
\includegraphics[width=14cm]{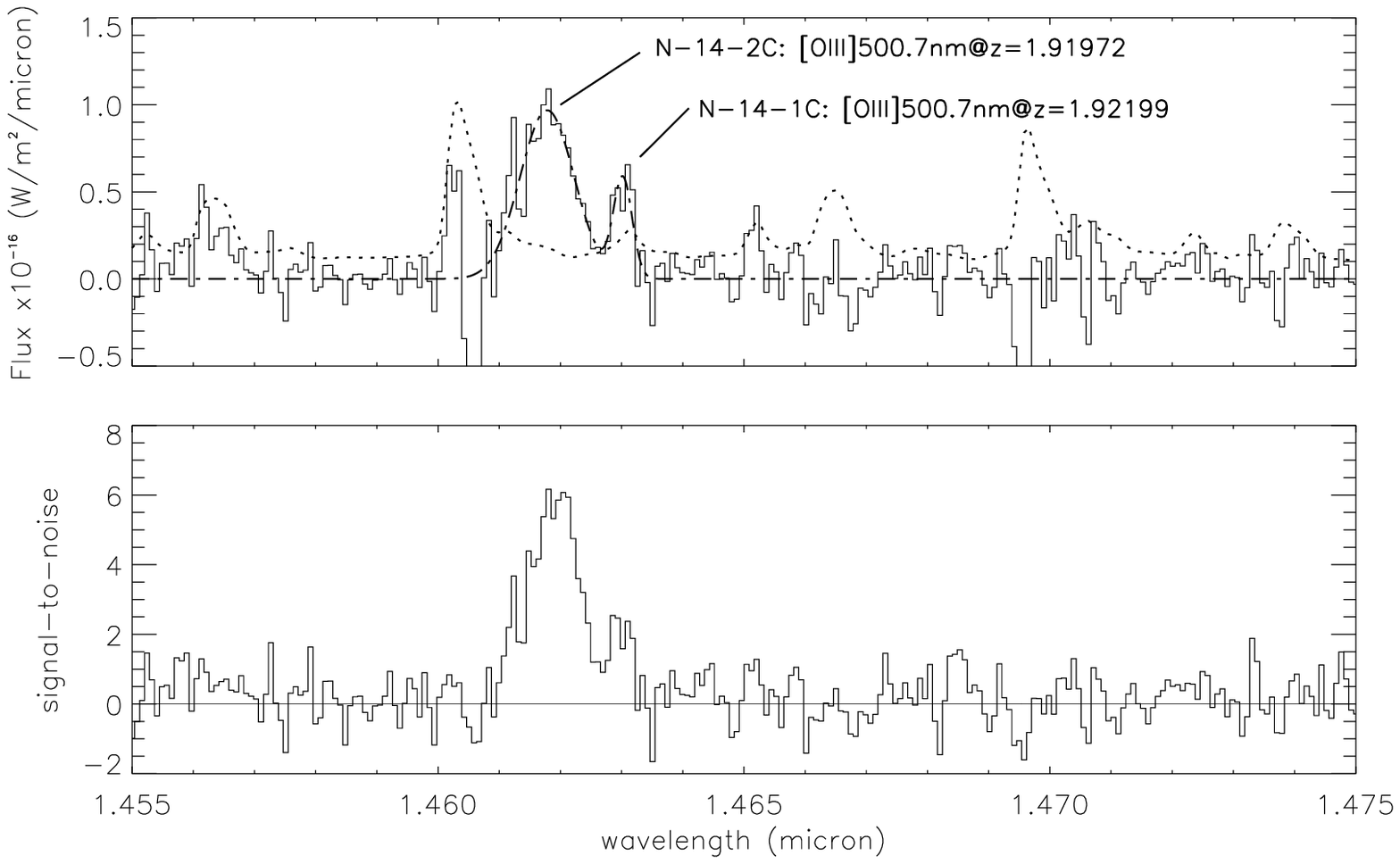}
\caption{ \label{fig:2206} One-dimensional extracted spectra of candidates N-14-1C and
N-14-2C. The top two panels show the data for N-14-1C, and the bottom
two panels show the data for N-14-2C. The top panel for each candidate
shows the extracted spectrum (solid), the 1$\sigma$ noise spectrum
(dotted), and the min--$\chi^2$ fit to the data (dot-dashed). For
N-14-1C we fit a single Gaussian to the emission line. For the
observation of N-14-2C we also detected light from N-14-1C, and we fit
two Gaussians, as described in the text. The bottom panel for each
candidate is the signal-to-noise spectrum. For N-14-1C this has been
binned up by a factor of four.}
\end{figure*}

\begin{figure*}
\includegraphics[width=14cm]{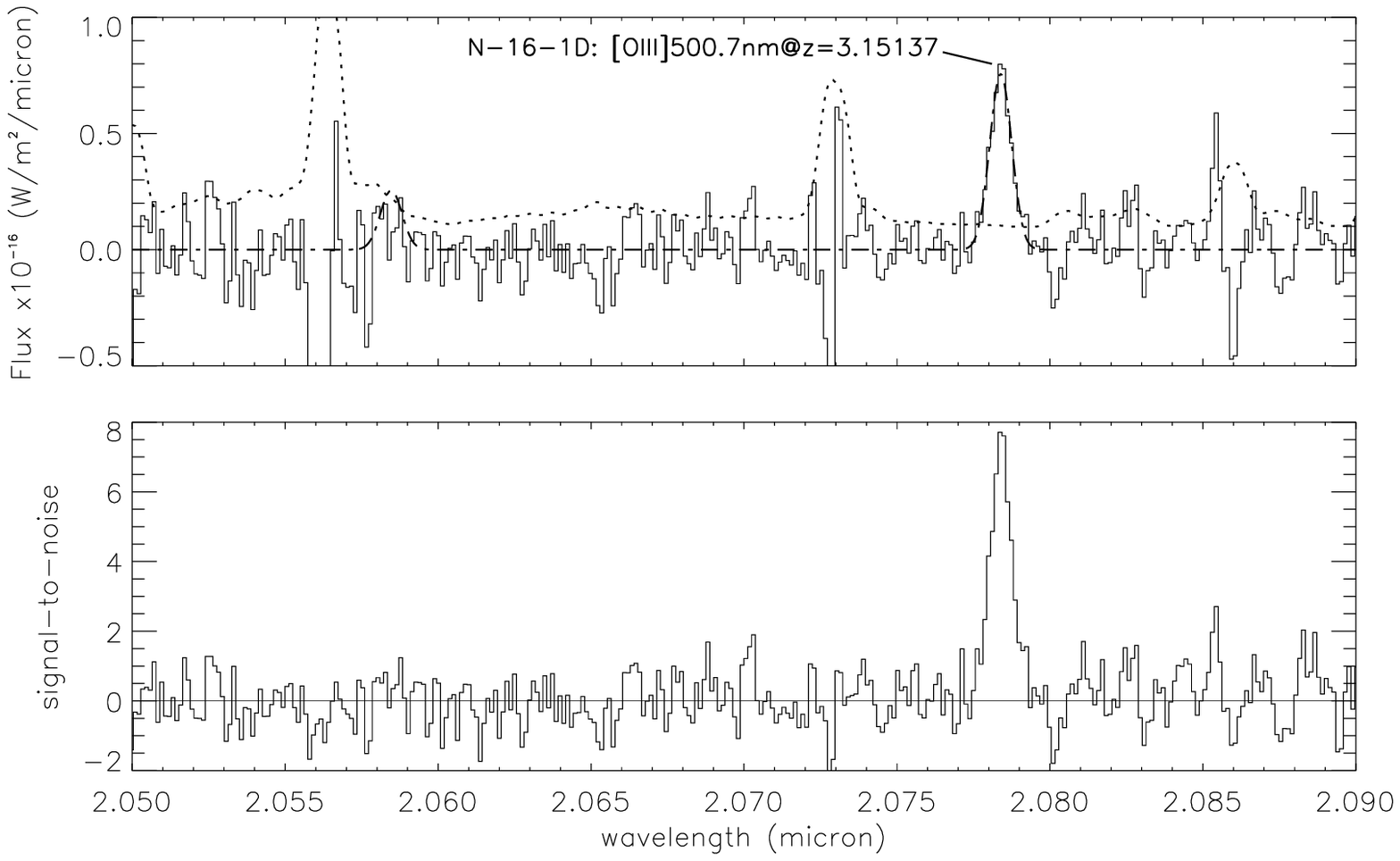}
\caption{\label{fig:2233}One-dimensional extracted spectra of candidate N-16-1D.  The
top panel shows the extracted spectrum (solid), the 1$\sigma$ noise
spectrum (dotted), and the min--$\chi^2$ fit of Gaussian emission
lines, fit simultaneously to the \OIII\,495.9, 500.7 lines
(dot-dashed). The flux from the weaker 495.9 line, near $2.059\mu$m,
is below the detection limit at that wavelength.  The bottom panel
shows the signal-to-noise spectrum.}
\end{figure*}

The galaxy 1D spectra are plotted in Figs \ref{fig:2206} and \ref{fig:2233}. In each case we
plot: i) the flux--calibrated 1D spectrum (solid), with the error
spectrum (dotted), and our min--$\chi^2$ fit of a Gaussian to the
detected emission line (dot--dashed), and ii) the ratio of the object
and error spectra i.e. the $S/N$ spectrum. The errors are large at the
wavelengths of the strong OH sky lines. Residuals from sky subtraction
can mean that the significance of a detected line in the plotted
spectrum may not always be easily appreciated. Plotting the $S/N$
spectrum can make this clearer.


Measurements of the three detected emission lines are summarised in
Table 4. Col. 1 lists the galaxy name, and col. 2 provides the
redshift of the DLA absorber (the meaning of `ELA' is explained in the
next section). For each detected emission line we made a min--$\chi^2$
fit of a Gaussian profile. The variables, then, are the line centre,
line width (corrected for resolution), and line flux.  Col. 3 provides
the redshift of the detected \OIII\ line; col. 4 lists the flux in the
\OIII\,500.7 line; and col. 5 lists the line luminosity. Col. 6
provides the intrinsic line width (FWHM), which is converted to a 1D
velocity dispersion in Col. 7. In Col. 8 we list the velocity
difference between the DLA absorption line and the \OIII\,500.7 emission
line. Col. 9 provides the velocity difference between any detected \lya\
emission and the \OIII\ emission.

\begin{table*}
\begin{center}
\begin{scriptsize}
\caption{\OIII\ emission line data for our three DLA galaxies}
\label{table:results1}

\begin{tabular}{lllcccccc}\\ 
\hline
 1 & 2 & 3 & 4 & 5 & 6 & 7 & 8 & 9  \\
Candidate & $z$DLA(ELA) & $z$([OIII]) & Flux $\times 10^{-17}$ & Lum $\times 10^{42}$ & FWHM &
 $\sigma$ & $\Delta$v(DLA-[OIII]) & $\Delta$v(Ly$\alpha$-[OIII])  \\
                 &                       &                         &
    ( \ergscm)            & (erg\thinspace s$^{-1}$) & (\kms) & (\kms) & (\kms) & (\kms) \\ 
    
    \hline
    
N-14-1C & \multirow{2}{*}{$1.919991(2)$ $^{1}$}  & $1.9220(2)$ & $(7.6\pm1.3)$  
& $(2.0\pm0.3)$ & $220\pm50$ & $90\pm20$ & $-210\pm20$ & $90\pm70$  \\

N-14-2C &      & $1.91972(8)$ & $(10.7\pm0.9)$ & $(2.8\pm0.2)$ & $180\pm25$ & $75\pm11$ 
& $29\pm9$ & ---  \\

N-16-1D & $3.14930(7)$ $^{2}$ & $3.15137(6)$ & $(6.78\pm0.5)$ & $(5.9\pm0.4)$ & $55^{+20}_{-30}$ 
& $23^{+8}_{-13}$ & $-150\pm6$ & $118\pm22$ \\

\hline
\end{tabular}
\begin{minipage}{170mm}

1. Component 2 from Table 5A in Prochaska and Wolfe (1997a)
2. Strongest FeII\,160.8 line (Lu, Sargent, and Barlow, 1997, 1998)

\end{minipage}
\medskip
\end{scriptsize}
\end{center}
\end{table*}

{\em Candidate N-14-1C:} The 1D spectrum of this candidate is shown in
the top two panels of Fig. \ref{fig:2206}. The \OIII\,500.7 line is detected at
$6\sigma$, at a redshift $z=1.9220$. The line has
FWHM$=220$km\,s$^{-1}$, corresponding to 14 pixels, which is greatly
oversampled. Therefore in panel b) we plot the $S/N$ in pixels binned
by a factor of four. In this plot the significance of the line is much
more apparent.

We checked for the possibility that the galaxy counterpart to the
$z=2.0762$ DLA absorber lies at small impact parameter, by searching
the spectrum of the quasar for the H$\beta$ and \OIII\,495.9 lines at
this redshift. The lines are undetected. Deeper limits were reached at
the other position angle, and are provided below.

{\em Candidate N-14-2C:} The 1D spectrum of this candidate is shown in
the lower two panels of Fig. \ref{fig:2206}. As stated previously, the \OIII\ line
from N-14-2C, detected at $S/N=12$ lies at a lower redshift than
N-14-1C. However the faint extension to the red, visible as a second
peak in the 1D spectrum, lies at a redshift consistent with the
redshift of N-14-1C. Therefore to measure the line we fit
simultaneously two Gaussians, with the higher--redshift line fixed at
the redshift of N-14-1C. As seen in the plot, this provides a good
fit. The weaker line is detected at $3.6\sigma$. There is no evidence
for light from N-14-2C leaking into the spectrum of N-14-1C, but given
the lower $S/N$ of the latter spectrum, and the slightly larger
distance of N-14-2C from the slit edge, Fig. 1, this is not
surprising.

We again checked for the possibility that the galaxy counterpart to
the $z=2.0762$ DLA absorber lies at small impact parameter, by
searching the spectrum of the quasar for the H$\beta$ and \OIII\,495.9
lines at this redshift. The lines are undetected. We measured fluxes
of $(0.2\pm4.4)\times 10^{-17}$\ergscm, and
$(-2.0\pm3.5)\times 10^{-17}$\ergscm for these lines,
respectively.

{\em Candidate N-16-1D:} The \OIII\,500.7 emission line is visible in
the two panels of Fig. \ref{fig:2233}. In measuring the properties of the line,
because the recorded spectrum covers both lines, \OIII\,495.9, 500.7,
we fit both lines simultaneously, with flux ratio 1:3. The \OIII\,500.7
line is detected at $S/N=13.5$. The \OIII\,495.9 line is not
significantly detected because it is a factor of three weaker, and
lies in a region of higher noise. The H$\beta$ line was not
detected. We measured a flux of $(1.1\pm1.6)\times 10^{-17}$ \ergscm for this line.



\section{Kinematics}

Prochaska and Wolfe (1997b, hereafter PW97)
have conjectured that DLA absorption lines arise in large, cold,
rapidly--rotating discs, of circular velocity $v_c\sim
225$km\,s$^{-1}$, from an analysis of the detailed velocity structure
of the low--ionisation metal absorption lines. Typically the strongest
component of the absorption complex occurs at either the blue or red
limit of the velocity spread, resulting in a characteristic asymmetric
profile with a sharp line edge. They were able to explain this
`edge--leading asymmetry' (hereafter ELA) in terms of the line of sight
passing through a thick disc. Consider a gaseous disc with a flat
rotation curve, of circular velocity $v_c$, and in which the number
density of absorbing clouds declines monotonically with distance from
the centre. The simplest case is if the galaxy is viewed edge--on,
$i=90$\thinspace degrees. Then the full circular velocity is recorded at the point
at which the line of sight passes closest to the centre. This also
corresponds to the strongest absorption if the face--on surface density
is a declining function of distance from the galaxy centre. Elsewhere
along the line of sight the column density is lower, and only a
component of the circular velocity is recorded. This explains the ELA
profile for the edge--on case. The line edge occurs at either the blue
or the red end of the profile, depending on whether the gas is
rotating towards or away from the observer, respectively.

At smaller inclinations, $i<90$\thinspace degrees, the situation is more
complicated, because the absorption profile depends also on the
density distribution of clouds perpendicular to the disc, and on the
point at which the line of sight crosses the disc mid--plane. In some
situations the absorption profile can be reversed, in the sense that
the strongest absorption can correspond to the minimum component of
rotational velocity (PW97, fig. 2). Finally, the absorption profile
also varies with impact parameter, in the sense that one observes a
smaller spread of velocities at larger impact parameter.

The PW97 picture of DLAs arising in large, cold, rapidly--rotating
thick discs is inconsistent with the CDM hierarchical model of
structure formation, which for these redshifts predicts smaller disks
with smaller rotation speeds. Nevertheless, as shown by Haehnelt et
al. (1998), the absorption line data itself may be equally well
explained by the CDM hierarchical model. In this case, the line
profiles are caused by a mixture of random motions, rotation, infall,
and mergers. Therefore the ELA test, on its own, cannot distinguish
between the two models.

As pointed out by Warren \& M\o ller (1996), PW97, and Lu, Sargent, \& Barlow
(1997, hereafter LSB), measurement of the galaxy systemic velocity,
through the detection of an emission line, provides a vital additional
datum for the interpretation of the kinematics and possible
distinction between these two pictures. The first application of this,
was the analysis by LSB of one of our targets, the $z=3.1501$ DLA
absorber in the spectrum of the quasar 2233.9$+$1381. LSB used the
redshift of the \lya\ emission line from N-16-1D, measured by
Djorgovski et al. (1996), as the galaxy systemic redshift; however,
the \lya\ line, being a resonance line, is subject to complex
radiation transfer processes, which result in an asymmetric emission
line profile. In the population of Lyman--break galaxies, Pettini et
al. (2001) find that, relative to the rest--frame optical emission
lines, the \lya\ emission line is redshifted by typically between 200
and 1000 \kms, while the interstellar absorption lines are blueshifted
by similar amounts. They interpret this as the signature of strong
galactic winds. The rest--frame optical emission lines, unaffected by
resonant scattering, provide the best measurement of the galaxy
systemic velocity, while the \lya\ line is unreliable.

We now turn to the interpretation of our results, in the light of this
discussion. Relevant quantities are summarised in Table 4. We discuss
the quasar 2233.9$+$1381 first, because it is the simpler case.

{\em Quasar 2233.9$+$1381:} The redshift of N-16-1D, from the
\OIII\,500.7 line, is $z=3.15137\pm0.00006$, which we take to be the
systemic redshift of the galaxy. The low--ionisation metal absorption
lines in this system display edge--leading asymmetry, towards the
blue. We are interested in the redshift of the ELA line edge, which we
take as recorded by the strongest Fe{\sc II}\,160.8 line, at
$z=3.1493$. This is entered in Table 4 as $zDLA(ELA)$. The absorption
line is plotted in fig. 1 of LSB, and the redshift is taken from Lu,
Sargent, \& Barlow (1998). No error is quoted, and we will adopt a
uncertainty of $\pm5$\kms, based on an inspection of the
spectrum. The DLA(ELA) line, then, is blueshifted by $150\pm6$\kms\
relative to the systemic velocity, while the \lya\ line is redshifted
by $118\pm22$\kms.

In the PW97 picture, these features (blue--side ELA line edge,
blueshifted relative to the systemic velocity, and a substantial
velocity difference) are the characteristics of a line of sight that
intersects the disc mid--plane close to the major axis, at small
impact parameter. An example is given by profile no. 3, in fig. 14 of
PW97. Because the intersection is close to the maor axis, we may
approximate the velocity difference of $150\pm6$\kms\ to $v_c
\sin{i}$. We note that this implies a substantially smaller mass than
inferred by LSB, who used the velocity difference given by the \lya\
line, $270$\kms, but is nevertheless in line with the proposal of PW97
of a typical circular velocity of $v_c\sim225$\kms.

Although the analysis presented so far sits well with the PW97
picture, a prediction of their model, not only for lines of sight such
as this, but for all lines of sight, is that the systemic velocity
should lie outside the velocity range of the low--ionisation metal
absorption--line profiles. This does not appear to be the case
here. LSB argue that the Fe{\sc II} line is the best tracer of the neutral
gas, and state that the velocity spread is $\Delta v\sim200$\kms, meaning
that the Fe{\sc II} absorption profile extends as far as $\sim50$\kms\
redward of the systemic velocity. This contradicts the prediction of
PW97. The velocity difference is significant, although not large, and
the effect would need to be seen in a number of systems before ruling out
the model.

{\em Quasar Q\,2206$-$1958:} A detailed analysis of the velocity
profiles of both low--ionisation and high--ionisation metal absorption
lines in the $z=1.9205$ absorber has been presented by Prochaska and
Wolfe (1997a). The velocity profile of the low--ionisation absorption
lines shows several components, with the two strongest separated by
65\kms.  There is a sharp cut--off to the blue of the lower--redshift
component (fig. 2 and table 5A in Prochaska and Wolfe, 1997a), and
they considered the profile ``to be consistent with an edge--leading
asymmetry, but note that its shape is not as suggestive as the
majority of the other cases''. Accepting this judgment, we select the
strongest blue component, $z=1.919991$, as the redshift of the leading
edge, entered as $z$DLA(ELA) in Table 4. The other components
identified extend redward over an interval of 140\kms.

We now have to choose between N-14-1C, which is redshifted by
$210\pm20$\kms\ relative to the ELA redshift, or N-14-2C which is
blueshifted by $29\pm9$\kms. The former is the simplest interpretation
within the PW97 picture. As for N-16-1D, above, the data are then
consistent with an interpretation of a line of sight that interesects
the mid--plane near the major axis, in which case we may approximate
the velocity difference of $210\pm20$\kms\ to $v_c \sin{i}$. Again this
value is in line with the proposal of PW97 of a typical circular
velocity of $v_c\sim225$\kms. Furthermore, in this case the systemic
velocity lies outside the velocity range of the low--ionisation metal
absorption--line profiles, in agreement with the prediction of
PW97. However, if the whole system is a large disk, it is then curious that N-14-1C is
fainter than N-14-2C, if it is the galaxy nucleus. A further
difficulty with this interpretation is the velocity of
N-14-2C. In the rotating disc hypothesis the only points blueshifted
by as much as the ELA line are other points along the major axis. But
this would pass from N-14-1C and run close to the quasar
position. N-14-2C lies almost perpendicular to this line (Fig. 1). Therefore
it must be concluded that it is a separate object. When we consider also
the disturbed appearance of the DLA galaxy in this
field, Fig. 1, these results are certainly more in agreement with the CDM
hierarchical picture than with the cold large disc hypothesis.

To summarise, the kinematic evidence from these two fields is in
agreement in a number of respects with the picture of PW97 of DLA
absorbers as arising in large rapdily--rotating cold
discs. Nevertheless there are two significant discrepancies with the
predictions of PW97: in the case of the quasar 2233.9$+$1381, the DLA
galaxy systemic velocity lies within the absorption velocity field,
rather than outside; in the case of the quasar Q\,2206$-$1958,
we are unable to incorporate the velocity of N-14-2C in a consistent
way, and this, coupled with the morphology of the system, supports the
hierarchical picture. Clearly the
detection of rest--frame optical emission lines from a larger sample of
DLA galaxies would be extremely useful, to extend this comparison.




\section{Comparison with the properties of LBGs}

\label{sec:LBG}

In this section we extend the analysis of M02, comparing the
properties of DLA galaxies and Lyman--break galaxies, to include
properties of the rest--frame optical emission lines. The aim is to
test the hypothesis that DLA galaxies are Lyman--break galaxies,
selected by gas cross section. As explained in M02, and in the
Introduction section here, although the average properties of the two
populations may differ (because of the different selection criteria),
if the hypothesis is correct, the properties of individual DLA
galaxies should lie within the range of the properties of a comparison
sample of Lyman--break galaxies of similar absolute magnitude and
redshift. This will be true regardless of how the DLA galaxies were
selected.

The new observations give us three extra parameters to use. These are
the luminosity weighted velocity dispersion, as recorded by the width
of the \OIII\ line, the \OIII\ line luminosity,
and the velocity difference between the \lya\ emission line and the
\OIII\ line.  Although our measurements of the three sources
considered here are for the \OIII\ line, we suppose that for the
comparison sample of LBGs any rest--frame optical emission line is
satisfactory for the first and third of these parameters.

The comparison data are drawn from the studies of Pettini et
al. (2001), and Erb et al. (2003), hereafter P01, E03, which provide
the largest samples of LBGs with spectroscopic detections of
rest--frame optical emission lines. P01 report observations of 19
targets, with mean redshift 3.1, and E03 report observations of 16
targets, with mean redshift 2.3. Some of the relevant quantities were
not included in these papers, but were kindly made available to us by
the authors. In total these two samples provide 32 measurements (or
upper limits) of the velocity dispersion, 11 measurements of the \OIII\
line luminosity, and 18 measurements of the velocity difference. For
consistency with M02, we use the AB continuum absolute magnitude at a
rest--frame wavelength of 150$\mu$m,, as the measure of galaxy
luminosity.

\subsection{Velocity dispersion}

\begin{figure}
\includegraphics[angle=90, width=1\columnwidth]{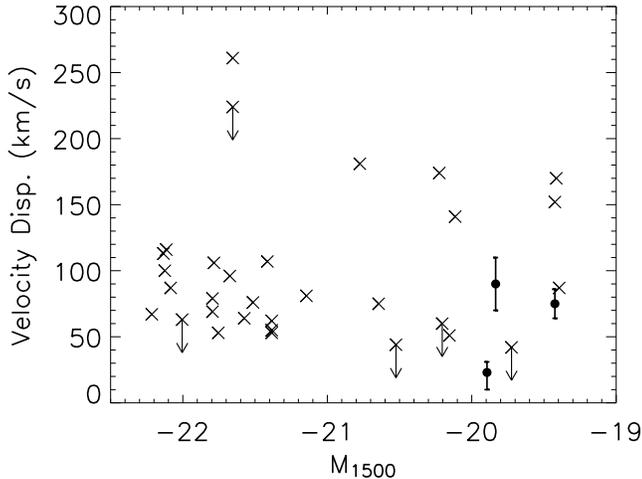}
\caption{ \label{fig:mag_disp} The central velocity dispersion of the rest--frame optical
emission line plotted against the 150nm absolute magnitude for the
sample of LBGs (crosses) and our DLA galaxies (filled circles).}
\end{figure}

In Fig. \ref{fig:mag_disp} we plot velocity dispersion against absolute magnitude for
our three sources, and the 32 LBGs. All values are intrinsic,
i.e. have been corrected for the resolution of the observations. Two
of the DLA measurements lie within the spread of values for the
LBGs. The third, N-16-1D, lies below the spread of
points. Nevertheless, a number of the LBG values are upper limits.We
conclude that there is no evidence for a difference between DLA
galaxies and LBGs, based on this plot.

\subsection{\OIII\ Luminosity}

\begin{figure}
\includegraphics[angle=90, width=1\columnwidth]{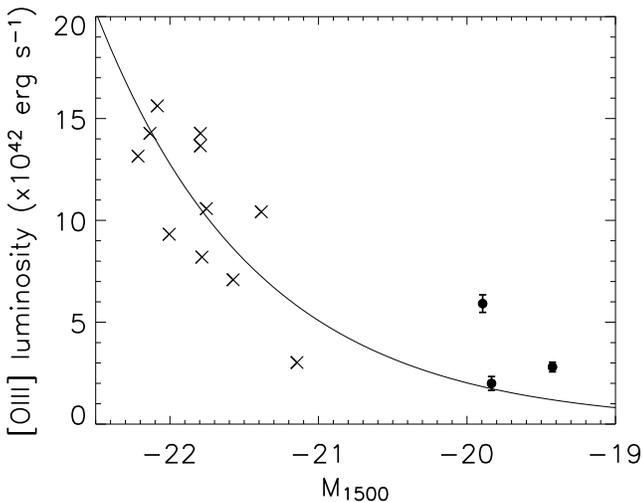}
\caption{ \label{fig:mag_l} The \OIII\ $\lambda$500.7 nm luminosity plotted against the
150nm absolute magnitude for the sample of LBGs (crosses) and our DLA
galaxies (filled circles).}
\end{figure}

In Fig. \ref{fig:mag_l} we plot \OIII500.7 line luminosity against absolute
magnitude for the three \OIII\ line detections reported here, and for the
11 LBGs. Unfortunately the two samples are disjoint in absolute
magnitude, preventing a valid comparison of the hypothesis under test.
The average \OIII\ luminosity of the DLA galaxies is a factor of three
smaller than the average luminosity of the LBGs. A trend of decreasing
line luminosity with rest--frame UV continuum luminosity would be
expected, since the continuum records the flux from the stars that are
the ionising sources. A linear relation between the two quantities, indicated
by the curve in Fig. \ref{fig:mag_l}, is a reasonably good representation of the trend.

\subsection{Velocity difference}

\begin{figure}
\includegraphics[angle=90,width=1\columnwidth]{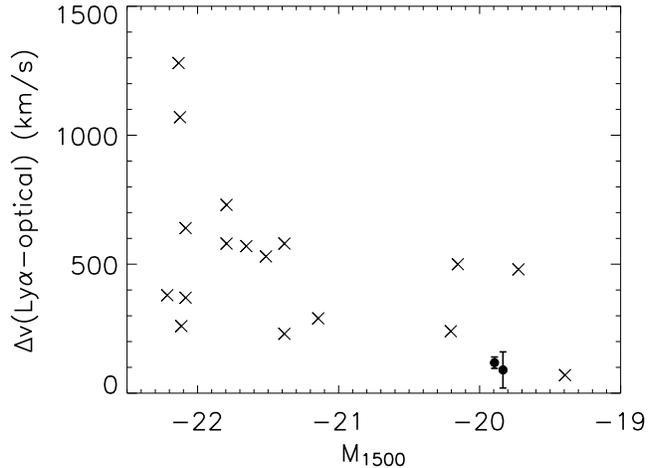}
\caption{\label{fig:delv}The velocity difference between the \lya\ emission line and
the rest--frame optical emission lines, plotted against the 150nm
absolute magnitude, for the sample of LBGs (crosses) and our DLA
galaxies (filled circles).}  
\end{figure}

For two of the three sources discussed in this paper, \lya\ emission
has been detected. The velocity differences between the \lya\
and \OIII\ lines, $\Delta v$, Table 4, are $90\pm70$\kms\ for N-14-1C,
and $116\pm22$\kms\ for N-16-1D. In Fig. \ref{fig:delv} we plot these values against
absolute magnitude, together with the data for the 18 LBGs. It is
striking that the values for the two DLA galaxies are smaller than all
the values for the LBGs, with the exception of one
galaxy. Nevertheless there is some evidence for a trend in this plot,
of smaller $\Delta v$ with fainter luminosity. Considering only the
small sample of four LBGs of similar luminosity to the DLA galaxies,
the two DLA galaxies lie within the range of the LBG galaxies, albeit
near the lower envelope. On this evidence we cannot rule out the
hypothesis of M02, and a larger sample of faint LBGs would be useful.

There are two possible selection effects which could explain the low
values of $\Delta v$ measured for the DLA galaxies. For DLA galaxies
at small impact parameter, \lya\ emission with large $\Delta v$ will
be much harder to detect, as lying in the wing of the absorption line,
rather than in the clean central region where the quasar flux is
completely absorbed. Secondly, any sample of DLA galaxies is liable to
be biased toward strong \lya\ emission, since this is how they are
confirmed spectroscopically. As shown by both Adelberger et al. (2003) and Shapley et al. (2003), in
LBGs self--absorption of the \lya\ line redshifts the emission line,
since the absorption occurs on the blue side. Therefore, galaxies
with weaker \lya\ emission tend to have larger velocity offsets, and
{\em vice versa}.

\section{Star formation rate}

It is possible to estimate the star formation rate (SFR) in these
three sources in a number of ways. We begin with the calibration of
Kennicutt (1998) relating SFR to unextinguished rest--frame UV
luminosity density:

\[
{\rm SFR}_{\rm FUV}(M_{\odot} {\rm yr}^{-1}) = 1.4 \times 10^{-28} \times L_{\rm FUV} ({\rm ergs \thinspace s}^{-1} \thinspace {\rm Hz}^{-1})
\]

The V band magnitudes of the three sources N-14-1C, N-14-2C, and
N-16-1D, are, respectively 24.69, 25.01, and 25.75. The first and
third of these values were taken from M02, while the second value has
not been published before. Photometry of the field of Q\,2206$-$1958
is quite difficult because of the irregular morphology of the
galaxy. The two values quoted here are aperture magnitudes within a
diameter of 0.6\thinspace arcsec. Without any correction for extinction these
yield the values of SFR quoted in Table 5, Col. 1. There is
substantial diffuse emission in the field of Q\,2206$-$1958. We have
estimated that the total flux in the field, from the two sources plus
the diffuse emission, corresponds to V=22.70. This value is
surprisingly bright, but is very uncertain because of the difficulty
of subtracting the bright image of the quasar. This total magnitude
corresponds to a SFR of 35\msunyr. Interestingly, Wolfe et al. (2004) 
find that the mean UV intensity in the absorber, measured from our STIS photometry,
is in good agreement with that inferred from spectroscopy of the C{\sc II}$^{\ast}$
absorption line -- supporting the use of this absorption line as an indicator of star formation.

The relation between SFR and unextinguished H$\alpha$ flux provided by
Kennicutt (1998) is:

\[
{\rm SFR}_{{\rm H}\alpha}(M_{\odot} {\rm yr}^{-1}) = 7.9 \times 10^{-42} \times L_{{\rm H}\alpha} ({\rm ergs \thinspace s}^{-1})
\]

Where available, H$\alpha$ (or H$\beta$) is preferred to the UV
continuum as a star formation gauge, because the observation is made
at a longer wavelength, and so is less affected by uncertain
corrections for extinction. Unfortunately H$\alpha$ is unobservable
from the ground for our two DLA absorbers, as is H$\beta$ for the
$z=1.92$ absorber towards Q\,2206$-$1958. For the $z=3.10$ absorber
towards 2233.9$+$1381 the H$\beta$ line, at $2.17\mu$m, lies in a
region of strong atmospheric absorption, and was not detected. Using
the $3\sigma$ upper limit to the line flux, and assuming
$\mathrm{H\alpha/H\beta=2.75}$ (Osterbrock, 1989), yields the upper
limit to the SFR for N-16-1D quoted in col. 2 of Table 5.

The \lya\ line is readily extinguished by dust, and therefore is only
useful in providing a lower limit to the SFR. Assuming
$\mathrm{Ly\alpha/H\alpha=10}$ yields the lower limits to the SFR
listed in col. 3 of Table 5, for the
two sources with detected \lya.

\begin{figure}
\includegraphics[angle=90,width=1\columnwidth]{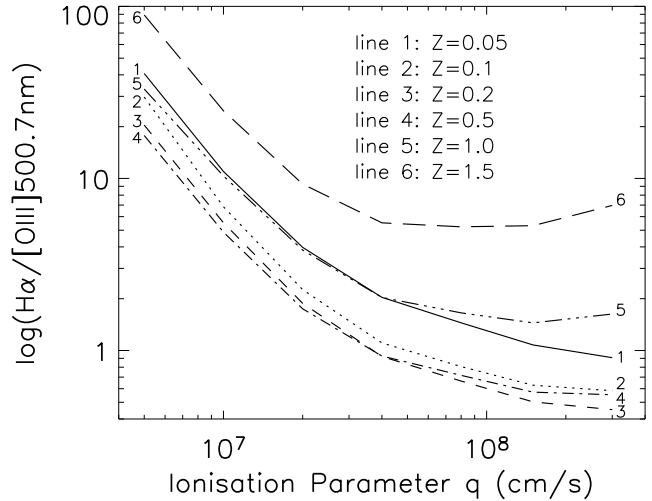}
\caption{ \label{fig:SFR} Ratio of \ha\ to \OIII500.7nm as a function of ionisation
 parameter, for different values of metallicity. The curves were
 derived by combining Figs 1 and 5 in Kewley \& Dopita (2002).}
\end{figure}

All the above estimates are subject to large uncertainty. Therefore it
is of interest to consider the usefulness of the \OIII\,500.7 line as a
star formation indicator. Kennicutt (1992) shows that there is large
scatter in a plot of \OIII\ luminosity against H$\alpha$ luminosity,
for a sample of nearby galaxies, and as a consequence advises against
its use as a star formation indicator. Differences in ionisation
parameter, $q$, and metallicity might explain the scatter.
To examine this further we used the
relations provided by Kewley \& Dopita (2002) to compute the
luminosity ratio H$\alpha$/\OIII\ as a function of these two
parameters. The results are plotted in Fig. \ref{fig:SFR}. This plot was derived by
combining the curves provided in figs 1 and 5 of Kewley \& Dopita
(2002), and assuming the ratio $\mathrm{H\alpha/H\beta=2.75}$ quoted
above.

\begin{figure}
 \includegraphics[angle=90,width=1\columnwidth]{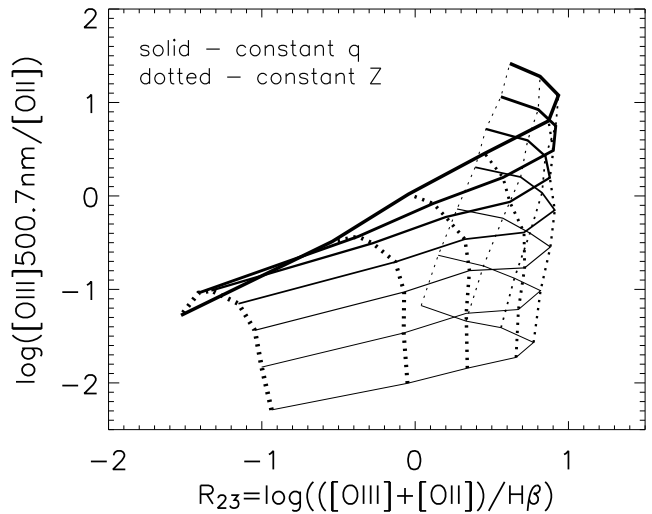}
 \caption{ \label{fig:SFR_diagnostic} Curves of constant ionisation parameter and constant
 metallicity as a function of \OIII/\OII and $R_{23}$. The curves were
 derived by combining Figs 1 and 5 in Kewley \& Dopita (2002). The solid lines
 shows curves of constant ionisation parameter, $q$: from thin to 
thick in the order $q$=[5e6, 1e7, 2e7, 4e7, 8e7, 1.5e8, 3e8].
 The dotted lines show curves of constant metallicity, Z: from 
thin to thick in the order Z=[0.05, 0.1, 0.2, 0.5, 1.0, 1.5, 2.0, 3.0]}
\end{figure}

Of particular interest here is the fact that within the range of
values of the metallicities for these two DLA absorbers, Table 1, the
ratio H$\alpha$/\OIII\ is very insensitive to metallicity, and near
the minimum value i.e. the inverse is near the maximum value. This
would explain why the \OIII\ lines were relatively easily detected in
these systems. For these metallicities the SFR depends much more
strongly on ionisation parameter than metallicity. It is then of
interest to estimate typical values of $q$ for LBGs. A useful diagram
for this is Fig. \ref{fig:SFR_diagnostic}. Here we have used the same two plots from Kewley
\& Dopita to derive curves of constant $q$ and constant metallicity as
a function of the two observables
$R_{23}=(\OII+\OIII)/\mathrm{H}\beta$ and $\OIII/\OII$. For values of
these two parameters for LBGs taken from table 3 of Pettini et
al. (2001), we determine typical values of $q\sim10^8$. Returning to
Fig. \ref{fig:SFR}, for the metallicities of the two DLAs, this implies that
H$\alpha$ is weaker than \OIII, with ratio H$\alpha/\OIII\sim0.6$.

To summarise this calculation: by using the measured metallicities of
the DLA absorbers; assuming that the ionisation parameters are
similar to those existing in LBGs; and assuming that the metallicities 
are applicable to the regions of star formation, we are able to estimate the
(unobservable) H$\alpha$ flux in these two DLA galaxies directly from
the \OIII 500.7 flux.  This leads to the star--formation rates entered
in col. 4 of Table 5.

We now compare the results between these various SFR indicators. For
N-14-1C, where \lya\ is strong, it is likely that there is not much
dust, and the fact that the \OIII\ SFR value is only a factor two
greater than the UV and \lya\ values agrees with this. The ratio of
the \OIII\ to UV estimate is larger for N-14-2C than for N-14-1C. This
would indicate that the system is dustier, which is in accord with the
redder colour and the absence of \lya\ emission. With the caveat that
the total magnitude for this field, including the diffuse emission, is
very uncertain, the \OIII\ estimates would suggest that the
correction for extinction to the total SFR estimate from the UV
continuum luminosity, given above, is about a factor of two; this indicates 
a total SFR, corrected for extinction, of about 70 $M_{\odot} {\rm yr}^{-1}$ for this system. Finally
the four estimates for N-16-1D are all in agreement with each
other. Nevertheless, in the context of the discussion for N-14-2C it
is surprising that the ratio of the \OIII\ to UV estimate is so large
for this object. Overall, there is gratifying agreement between the
different methods used to estimate SFRs.

Returning now to the use of the \OIII\ luminosity as an indicator of
the star formation rate, the foregoing analysis would suggest that,
provided independent information on the ionisation parameter and
metallicity are available, the \OIII\ luminosity can provide an
estimate of the star formation rate that is accurate to better than a
factor of two. Clearly this is not a method that has any general
application, but it could be useful for other DLA galaxies, where \OIII\
may be the only detectable line.

\begin{table}
\begin{center}
\begin{scriptsize}
\caption{Star formation rates for the three DLA galaxies, $M_\odot$yr$^{-1}$}
\label{table:SFR}

\begin{tabular}{lcccc}\\ 
\hline
Candidate & UV & H$\beta$ & Ly$\alpha$ & \OIII \\

\hline
N-14-1C & $5.7$ & --- & $>5.6$ & $9.5$ \\
N-14-2C & $4.2$ & --- & --- & $13.3$ \\
N-16-1D & $5.9$ & $<112$\thinspace $^\ast$ & $>4.4$ & $28$ \\
\hline
\end{tabular}
\begin{minipage}{60mm}
$^\ast$ $3\sigma$ upper limit
\end{minipage}

\medskip
\end{scriptsize}
\end{center}
\end{table}



\section{Summary}

\label{sec:summary}

We summarise the main results of the paper.

1. We have detected \OIII\ 500.7 emission lines from three sources
associated with high--redshift $z>1.75$ DLA absorption lines. These
are the first detections of their kind. Two of the sources are
identified with the same DLA absorber, at $z=1.92$, seen in the
spectrum of the quasar Q\,2206$-$1958, and the third source is the
galaxy counterpart to the $z=3.10$ absorber seen in the spectrum of
the quasar 2233.9$+$1381, previously detected in \lya.

2. Comparison of the \OIII\ emission redshift with the detailed
velocity structure seen in the absorption profile of low--ionisation
metal lines, allows a new test of the model of Prochaska and Wolfe
(1997b) of DLA absorbers as large rapidly--rotating cold thick
disks, which contradicts the CDM hierarchical picture. 
Some of the predictions of this model are borne out by the
data, but we note two significant discrepancies. The detection of
rest--frame optical emission lines from DLA absorbers provides a
valuable test of this model. More detections could provide a definitive test.

3. The properties of the detected rest--frame optical emission lines
provide further investigation of the question posed by M\o ller et al. (2002),:
are DLA galaxies are Lyman--break galaxies selected by gas cross
section? Comparison of the measured velocity dispersions, \OIII\ line
luminosities, and the velocity difference between the \OIII\ and \lya\
lines reveal no clear contradictions of the hypothesis. Again a larger
sample would be valuable in testing this suggestion more strongly.

4. We have described a method to estimate the SFRs in these two DLA
 galaxies using the \OIII\ luminosity, and assuming that the measured
 metallicity of the DLA is applicable to the regions of star
 formation. We find gratifying agreement between the SFRs estimated in
 this way, and those estimated using the traditional indicators.



\section*{Acknowledgments}

We thank Max Pettini, Dawn Erb and Alice Shapley for providing further
data and useful communications on the Lyman--break galaxy sample.



\bsp

\label{lastpage}

\end{document}